\documentclass[aps,pra,amsmath,amssymb,amsfonts,superscriptaddress,floatfix]{revtex4} 
\pdfoutput=1
\usepackage{amsmath,amsthm,amsfonts,amssymb}
\usepackage{graphicx}
\usepackage{natbib,hyperref}
\usepackage{doi}

\usepackage{color}

\newtheorem{theorem}{Theorem}

\newtheorem{lemma}{Lemma}

\newtheorem{conjecture}{Conjecture}

\usepackage{algorithm}
\DeclareMathAlphabet{\mathpzc}{OT1}{pzc}{m}{it}

\newcommand{\be}{\begin{equation}}
\newcommand{\ee}{\end{equation}}

\newcommand{\mO}{{\mathcal O}}
\newcommand{\pplus}{\psi_{+}}
\newcommand{\expec}{\mathbb{E}}
\newcommand{\pr}{{\rm Pr}}

\begin{document}
\bibliographystyle{plainnat}

\title{Duality in Quantum Quenches and Classical Approximation Algorithms: Pretty Good or Very Bad}

\author{Matthew B.~Hastings}

\affiliation{Station Q, Microsoft Research, Santa Barbara, CA 93106-6105, USA}
\affiliation{Quantum Architectures and Computation Group, Microsoft Research, Redmond, WA 98052, USA}
\begin{abstract}
We consider classical and quantum algorithms which have a duality property:
roughly, either the algorithm provides some nontrivial improvement over random or there exist many solutions which are significantly worse than random.  This enables one to give guarantees that the algorithm will find such a nontrivial improvement: if few solutions exist which are much worse than random, then a nontrivial improvement is guaranteed.
The quantum algorithm is based on a sudden {\it quench} of a Hamiltonian;
while the algorithm is general, we analyze it in the specific context of MAX-$K$-LIN$2$, for both even and odd $K$.
The classical algorithm is a ``dequantization of this algorithm", obtaining the same guarantee (indeed, some results which are only conjectured in the quantum case can be proven here); however, the quantum point of view helps in analyzing the performance of the classical algorithm and might in some cases perform better.
\end{abstract}
\maketitle

\section{Introduction}
For many combinatorial optimization problems, we expect that it is not possible to obtain an exact solution in polynomial time.  Instead, the best
that we can hope for is to obtain an {\it approximate} solution.  
The main result of this paper is a duality for certain approximations, that one can call ``pretty good or very bad", in which the algorithm either finds a nontrivial improvement over random (``pretty good") or there exist many solutions which are significantly worse (``very bad").  
This can lead to a method of proving guarantees of performance for the algorithm, if one knows that such very bad solutions do not exist.

We were led to this duality by analyzing a quantum approximation algorithm based on the idea of a {\it quench}: a sudden change in the Hamiltonian; more specifically, we prepare the system in the ground state of a given Hamiltonian, and then evolve it under a different Hamiltonian.  We present this algorithm in this paper and give some evidence for the duality there.  We then
consider a classical approximation algorithm and prove the duality there.  Finally, we discuss the performance of the quantum algorithm on certain instances.

We will consider the optimization problem MAX-$K$-LIN-$2$, with the assumption of a degree bound explained below.
Roughly speaking, this problem MAX-$K$-LIN-$2$ considers an objective function which is a sum of terms of order $K$ in binary variables; we give
a more precise definition later (we use the term ``order" rather than ``degree" to denote the exponent of a polynomial to avoid confusion with the use of ``degree" for the degree bound).  We will call these binary variables ``bits", though we emphasize that they take values in $\{-1,+1\}$ rather than $\{0,1\}$.

We consider an instance with degree $D$, so that each bit participates in $D$ terms in the problem.
Previous work has shown that for odd $K$ it is possible to obtain a nontrivial approximation of order $1/\sqrt{D}$ for MAX-$K$-LIN-$2$ using a classical algorithm\cite{barak2015beating} (initially a quantum algorithm was found\cite{farhi6062quantum} providing weaker approximation guarantees but later the classical algorithm was discovered).
Further, for arbitrary $K$ the classical algorithm finds a solution which is either better than random by an amount $1/\sqrt{D}$ or worse by an amount of order $1/\sqrt{D}$.  This result implies the order $1/\sqrt{D}$ improvement for $K$ odd, since if the algorithm finds a result which is worse than random by order $1/\sqrt{D}$, one can change the sign of all bits to obtain an improvement by order $1/\sqrt{D}$.

We consider a different but closely related classical approximation algorithm and find (for arbitrary $K$, though the result is most interesting for even $K$) the duality mentioned above which generalizes this: rather than being better or worse by $1/\sqrt{D}$, one can instead choose it to be slightly better or much worse.  There is a constant $\epsilon$ one may choose and roughly (precise results are in theorem \ref{dualclemma} below) the algorithm either finds a solution which improves on random by an amount $\epsilon/\sqrt{D}$ or there is a solution which is worse than random by an amount $\epsilon^{-1}/\sqrt{D}$.
For example, if one chooses $\epsilon$ slightly larger than $1/\sqrt{D}$, the algorithm either improves on random by an amount more than $1/D$ (a ``pretty good" solution) or there exists a solution which is worse than random by almost $1$ (a ``very bad" solution).
The improvement by more than $1/D$ is important because
it is always possible to find an assignment which improves by a factor $1/D$ in 
polynomial expected time\cite{haastad2000bounded},  i.e, such an improvement can be found in polynomial expected time regardless of the value of the optimal assignment.

We also analyze a quantum algorithm based on quenches.  Rather than slowly changing a Hamiltonian as in the adiabatic algorithm\cite{farhi2001quantum} (which in general is expected to have trouble with small energy gaps\cite{altshuler2010anderson}), we suddenly change the Hamiltonian, but then spend some time evolving under the new Hamiltonian.  We propose this algorithm as a general method for approximate optimization, but we analyze it in the context of MAX-$K$-LIN-$2$.  Here we find a similar duality.

The quantum algorithm gives a point of view that is useful in analyzing the classical algorithm: both algorithms find improvements unless there is a quantum state with large polarization in the $X$ direction (i.e., the expectation value of the sum of Pauli $X$ operators on all qubits is large as defined below) and which has an expectation value for the objective function which is significantly worse than random.  Some of the results in the quantum case are only conjectured, while they can be proven in the classical case.  However, the
quantum algorithm may be useful for some other instances.
  
  \subsection{Problem Definition and Examples}
 We consider the problem MAX-$K$-LIN-$2$.  There are $N$ variables, called bits, each of which may take values in $\{-1,+1\}$.  The objective function, which we denote $H_Z$, is taken to be a weighted sum of monomials of
order $K$ in these bits, i.e., each monomial is a product of $K$ distinct bits (sometimes this problem is called MAX-E$K$-LIN-2 to distinguish it from a more general case where monomials may have order up to $K$).  We will require that the weight of each monomial be chosen from $\{-1,+1\}$, and that all monomials be distinct from each other.

We consider an optimization problem where the goal is to maximize this objective function.  We emphasize this because we will later consider a Hamiltonian which includes a term proportional to $H_Z$, and so we will be considering states near the highest energy state of that Hamiltonian, rather than the lowest energy state as more commonly done in physics.

We write the bits as $Z_i$ where $i\in \{1\,\ldots,N\}$ so that there are $N$ bits, so for MAX-$2$-LIN-$2$ we have
\be
H_Z=\sum_{i<j} J_{ij} Z_i Z_j,
\ee
where $J_{ij}$ is a matrix with entries chosen from $\{-1,0,+1\}$.

We will assume a degree bound $D$, so that each bit $Z_i$ appears in at most $D$ distinct monomials in $H_Z$.  Indeed, for simplicity
we will only consider the case where each $Z_i$ appears in exactly $D$ monomials in $H_Z$.
We define $N_T$ to equal the number of terms in $H_Z$ so that if every bit has degree exactly $D$ and every term is order exactly $K$ then we have
\be
N_T=\frac{D N}{K}.
\ee

A random assignment has expectation value of $H_Z$ equal to $0$.  
Typically in computer science, one regards each of these monomials as a {\it constraint}: the constraint is satisfied if the monomial is equal to $+1$ and it is violated otherwise, so that the number of satisfied constraints is equal to the value of $H_Z/2$ plus $N_T/2$.  Hence, a random assignment satisfies half the constraints on average.  Then, the approximation ratio achieved by some assignment to the bits is defined to be the fraction of constraints satisfied by that assignment divided by the fraction of constraints satisfied by the optimal assignment.

We will define the approximation ratio differently: we will define it to be the value of $H_Z$ for a given assignment divided by the value of $H_Z$ in the optimal assignment.  That is, we will not add this term $N_T/2$.

We will also say that an assignment improves by a factor $f$ over random if it has $H_Z\geq f N_T$.
We say that an assignment is worse than random by a factor $f$ if it outputs an assignment $H_Z\leq -fN_T$.

For odd $K$, it is possible to improve over a random assignment by $\exp(-O(K))/\sqrt{D}$ in polynomial expected time\cite{barak2015beating}.
One cannot expect to have such an improvement for even $K$ simply because there exist families of instances in which no assignment
has $H_Z$ larger than $N_T \cdot \mO(1/D)$.
For $K=2$, a simple such example is to choose
\be
\label{ex2}
H_Z=-\sum_{i<j} Z_i Z_j.
\ee
Here we have taken $D=N-1$ so that every bit is in some monomial with another bit.
It is possible to obtain a very large negative expectation value of $H_Z$ (i.e., $-N(N-1)/2$) by choosing all $Z_i$ to have the same sign, but for even $N$, the maximum positive expectation value of $H_Z$ is to choose $N/2$ of the $Z_i$ to equal $+1$ and the remainder to equal $-1$, giving expectation value $N/2$, which is proportional to $N_T/N$.
This example provides an early example of the duality: the maximum improvement over random is quite small (a factor $O(1/D)$) but one can find an assignment which is a factor $\Omega(1)$ worse than random.

For $K=2m$, one can generalize example (\ref{ex2}) to give an instance for MAX-$K$-LIN-$2$ as follows: let $N=mD$.  
Divide the set of $mD$ bits into $D$ disjoint sets, each containing $m$ bits.
Label the sets by integers in ${1,\ldots,D}$.  Let $\tilde Z_i$ be the product of the bits in the $i$-th set.
Let $H_Z=-\sum_{i<j} \tilde Z_i \tilde Z_j$.

\subsection{Outline}
In section \ref{qa} we define the quench algorithm, both in the specific form that we analyze later as well as some variants that may be useful.
Subsection \ref{qdualsec} shows how the duality arises in the quantum algorithm; here we need to make some conjectures to show that the duality
holds.
In section \ref{rounding} we collect some results that will be useful in analyzing the classical algorithm that we give later as well as in analyzing the quantum algorithm.
In section \ref{ca} we define the classical algorithm and analyze it; in contrast to the quantum case, we will be able to prove all the conjectured results about the classical algorithm.
In section \ref{mf} we consider some applications of the analysis of these algorithms.
In section \ref{aqa} we give a further analysis of the quantum algorithm in an attempt to support the conjectures of subsection \ref{qdualsec}.

\section{Quench Algorithm}
\label{qa}
To define the algorithm, we promote the bits to qubits, and we let $Z_i$ be the Pauli $Z$ operator on the $i$-th qubit.
Let $X_i$ be the Pauli $X$ operator on the $i$-th qubit and let
\be
X=\sum_i X_i.
\ee

We use the following algorithm.
Let
\be
H=X+\frac{\alpha}{D} H_Z,
\ee
where $\alpha$ is a scalar chosen later.
We prepare the system in the state $\pplus$ maximally polarized in the $+$ direction so that the expectation value of $X_i$ is equal to $+1$ for all $i$.
We then evolve the system under Hamiltonian $H$ for a time $T$ that we choose later.  This time will in all case be at most ${\rm poly}(N)$; indeed, our analysis will be for $T=O(1)$.
Hence, this evolution can be performed in polynomial time on a quantum computer in time polynomial in $t_{max}$ and polynomial in the inverse error using any of a number of algorithms\cite{QSP,LC16,BerryEtAl2014,TS,BCK15} 
 (indeed, the simulation can be performed in time polylogarithmic in the inverse error for some of these but we will not need this).
In any simulation algorithm on a quantum computer, we discretize the variable $t$; for example, one may choose it to equal
an integer multiple of some time $t_{min}$ for some $t_{min}$ which is polynomially small; this causes only a polynomially small error.
Finally, we measure the state of the system in the computational basis, giving an assignment of bits $Z_i$.

This algorithm can be regarded as an example of a quantum walk on a hypercube\cite{childs2004spatial,qrev}.  While the present paper was in preparation, another paper used these quantum walk ideas to give a closely related algorithm which was then analyzed numerically in the context of the Sherrington-Kirkpatrick and random energy models\cite{callison2019finding}.

In the analysis of the algorithm, we will ignore all the errors associated with the time evolution and the discretization of time, since a polynomially small error is negligible as may be verified.

When we apply this algorithm, one may repeat the algorithm several times with $T$ chosen from an appropriate distribution.
In this regard, it is interesting to think about the state arising from averaging $T$ over an interval of times; by choosing the time from a random distribution (or alternatively by performing phase estimation of the Hamiltonian $H$) we can decohere the system in an eigenbasis.
The fixed evolution has a similar effect but is easier to analyze using the techniques here.
We can also use a similar idea to that in Ref.~\cite{Poulin} and simulate a {\it function} of the Hamiltonian which should have a similar effect but may be faster to simulate.

\subsection{Motivation}
Let us heuristically explain the algorithm.
The time evolution has two purposes.  The first is to decohere different eigenstates of the Hamiltonian as mentioned; for fixed time, the evolution for time $t$ produces a pure state, but produces some change in phase for different energies which has a similar effect to a random evolution.  The second purpose is to do it in a way that conserves energy.  
One hopes that the decoherence between different eigenstates will lead to a reduction in the expectation value of $X$, since one hopes that individual eigenstates will not have large $X$.
This reduction will lead to a positive expectation value of $H_Z$ due to the energy conservation as we now explain: this energy conservation is the second reason for the time evolution.

For arbitrary operators $O,H$ and scalar $t$,
define $\tau^H_t(O)=\exp(i t H) O \exp(-i t H)$.
Define
\be
\langle O \rangle_+\equiv \langle \pplus | O | \pplus \rangle.
\ee
Define
\be
\langle O \rangle_T\equiv \langle \tau^H_T(O) \rangle_+,
\ee
so that $\langle O \rangle_T$ is the expectation value at time $T$.
We have
\be
\langle H \rangle_T = \langle H \rangle_+=N,
\ee
independent of $T$
by energy conservation. Hence, we have
\be
\label{energybal}
\langle H_Z \rangle_T =D \frac{N-\langle X\rangle_T}{\alpha}.
\ee
That is, if the state at time $T$ has an expectation value of $X$ that is smaller than the maximal (i.e., smaller than $N$), it necessarily has an expectation value of $H_Z$ that is positive, i.e., it has obtained some solution that is better than random.
This is the key idea behind the quench algorithm.

Note that if the algorithm obtains a state with a large expectation value of $H_Z$ (much larger than $N_T/D$), then since the expected value is within $1/{\rm poly}(D)$ of the optimal value (which is at most $O(N_T)$), by repeating the algorithm ${\rm poly}(D)$ times we can, with probability at least one-half, obtain a solution which is at least a constant factor
times
the expected value.
 Here the constant factor can be any number strictly less than $1$, for example $1/2$.
This is an application of Markov's inequality.  Consider $\Vert H_Z \Vert - H_Z$ as a non-negative random variable with expectation value $\Vert H_Z \Vert - \expec[H_Z]$, where $\Vert H_Z \Vert=O(N_T)$ denotes the largest eigenvalue of $H_Z$ in absolute value.  The probability that $H_Z$ is smaller than $\expec[H_Z]/2$, for example, is bounded by $(\Vert H_Z \Vert-\expec[H_Z])/(\Vert H_Z \Vert-\expec[H_Z]/2)=1-\Omega(1/{\rm poly}(D))$ so with probability $\Omega(1/{\rm poly}(D))$ we have that $H_Z \geq \expec[H_Z]/2$.

\subsection{Heuristic Choices of $\alpha$ }
\label{Calpha}
We now discuss how to choose $\alpha$.  We give a calculation that introduces some of the notation used later. 
We consider perturbation theory to only second order,
and we then give a purely heuristic treatment of higher orders to motivate the choice of $\alpha$.  Later we will give a different treatment.

Consider the series for $\tau^H_T(X_i)$ for some given $i$:
For any operator $O$, we have the series
\begin{eqnarray}
\label{ho}
\tau^H_T(O)=O-iT [O,H]-\frac{T^2}{2}[[O,H],H] +i\frac{T^3}{3!}[[[O,H],H],H] + \frac{T^4}{4!}[[[[O,H],H],H],H]+\ldots
\end{eqnarray}

So, we have
\be
\label{tauTX}
\tau^H_T(X)=X-i\frac{\alpha T}{D}[X,H_Z]-\frac{\alpha}{D}\frac{T^2}{2} [[X,H_Z],H]+\ldots,
\ee
where the dots denote terms of order $T^3$ or higher.
Hence,
\begin{eqnarray}
\label{Xo2}
\langle X\rangle_T &=& \langle X \rangle-\frac{\alpha^2}{D^2} \frac{T^2}{2}[[X,H_Z],H_Z] \rangle_+ + \ldots \\ \nonumber
&=& N-2\frac{\alpha^2}{D}T^2 N + \ldots,
\end{eqnarray}
where we use that $\langle [[X,H_Z],H] \rangle_+=\frac{\alpha}{D}\langle [[X,H_Z],H_Z] \rangle_+=4N\alpha$.
and so by Eq.~(\ref{energybal}),
\be
\langle H_Z \rangle_T = 2\alpha T^2 N+\ldots
\ee

Of course, the higher order corrections to this perturbation theory must become important for large enough $T,\alpha$.
For one thing, once $T\gtrsim 1$, the effects of higher order terms in $TX$ in the exponential become important, i.e., we must consider higher order commutators such as $[[[[X,H_Z],X],X],H_Z]$.
However, we might hope that for some $T$ of order unity (for example, $T=1/2$) the  higher orders in $T$ will not be too important; maybe they will not be negligible but we might hope that they will only slightly reduce the result.

However, even for such a fixed $T=1/2$, we certainly cannot ignore higher order terms in $(\alpha/D) H_Z$ for large enough $\alpha$.
For example, if $\alpha$ is sufficiently larger than $\sqrt{D}$, we would find that Eq.~(\ref{Xo2}) gives a result for $X$ which
is smaller than $-N$, which is impossible.

So, the most optimistic thing we can hope for is that
second order perturbation theory is roughly accurate up to some $T$ of order unity such as $T=1/2$ and up to $\alpha$ proportional to $\sqrt{D}$.
If so,
we would find that the best choice of $\alpha$ would be to take $\alpha$ proportional to $\sqrt{D}$, in which
case we would have $\langle H_Z \rangle$ proportional to $N \sqrt{D}$, which is proportional to $N_T \cdot \Omega(1/\sqrt{D})$.  Thus it would give an $\Omega(1/\sqrt{D})$ factor approximation.

However, clearly this heuristic analysis is too optimistic.  Such solutions do exist for MAX-$K$-LIN-$2$ for odd $K$ (though we certainly have {\it not} shown that the algorithm finds them), but they do not exist in general, such as the example of Eq.~(\ref{ex2}).

\subsection{Duality}
\label{qdualsec}
The previous subsection considered a perturbative approach; the second order term corresponded to considering the second derivative with respect to $T$
of $\langle X \rangle_T$ at $T=0$.
We now consider this derivative at arbitrary $T$.

We introduce some notation that will be useful both here and later, including for the classical algorithm.
Let us define $F_i$ (the symbol ``F" is for ``force", i.e., a derivative of energy with respect to some coordinate) to equal $Z_i$ times the sum of terms in $H_Z$ that include $Z_i$.  For example, for $K=4$ and $H_Z=Z_1 Z_2 Z_3 Z_4 + Z_1 Z_3 Z_4 Z_5$ then $F_1=Z_2 Z_2 Z_4+Z_3 Z_4 Z_5$.  Note that the multiplication by $Z_i$ reduces the order of the terms in $F_i$ to $K-1$ since $Z_i^2=1$.  The ``force" depends upon the choice of $Z_i$ so we will sometimes write $F_i(\vec Z)$ to indicate its dependence on $\vec Z$.

Considering this second derivative at arbitrary $T$ we have
\be
\partial_T^2 \langle X_i \rangle_T=-4\frac{\alpha^2}{D^2} \langle X_i F_i^2 \rangle_T+2\frac{\alpha}{D} \langle 2 Z_i F_i - Y_i \dot F_i \rangle_T,
\ee
where
for any operator $O$, we define $\dot O=-i [O,H]$.

For $T=0$ the first term is equal to $-4\frac{\alpha^2}{D}$.  Assuming (we consider this in more detail later) that the first term remains $-\Omega(1) \frac{\alpha^2}{D}$ for $T\neq 0$, then we have
$\langle X_i \rangle_T=1-\Omega(1) \frac{\alpha^2 T^2}{D}$ {\it unless} the second term 
$\frac{\alpha}{D} \langle  2 Zi F_i -Y_i \dot F_i \rangle_T$ also becomes $\Omega(1) \frac{\alpha^2}{D}$.
For this to happen, we need 
$ \langle \tau^H_T(2Z_i F_i -Y_i \dot F_i) \rangle_+=\Omega(1) \alpha$.

Heuristically speaking, and ignoring the correlation between $X_i$ and $F_i^2$, one way that the first term could become small is for the expectation value of $X_i$ to become small.  This would of course mean a state with large expectation value of $H_Z$.  Another way is for $F_i^2$ to become small.

Thus, under the assumption about the first term, we have at least one of two situations.
Either, after time $T$, we have 
$\langle X \rangle_T=N\cdot(1-\Omega(1) \frac{\alpha^2 T^2}{D})$ so that
$$\langle H_Z \rangle_T=\Omega(1) \cdot \alpha T^2 N$$
or
we have 
$$\sum_i \langle 2Z_i F_i - Y_i \dot F_i \rangle_s=\Omega(1) \alpha N$$ for some time $s\leq T$ (or both possibilities may hold).  Further, at that time $s$, if we do not have $\langle X\rangle_T \geq N\cdot(1-O(1) \frac{\alpha^2 T^2}{D})$ then we have
$\langle H_Z \rangle_T=\Omega(1) \alpha T^2 N$.

So, we conjecture:
\begin{conjecture}
\label{conjer}
For $\alpha^2 T^2 =O(1)$, if the algorithm does not find a state (by sampling over times $s\leq T$) with expectation value of $H_Z$ equal to $\Omega_K(1) \alpha T^2 N$, then there exists some state with expectation value of $X$ at least $N\cdot(1-O(1) \frac{\alpha^2 T^2}{D})$
and expectation value of $\sum _i (2 Z_i F_i - Y_i \dot F_i)$ at least $\Omega_K(1) \alpha N$.
\end{conjecture}
Here the notation $\Omega_K(\ldots)$ denotes that the constant may depend upon $K$ but not on $\alpha,T,N,D$.

Choosing $\alpha^2 T^2\sim D$, we see that either the algorithm finds a solution with
expectation value of $H_Z$ equal to at least $$\Omega_K(\frac{D}{\alpha}) N$$
or there exists some state with
expectation value of $\sum _i (2 Z_i F_i - Y_i \dot F_i)$ at least $$\Omega_K(\alpha N),$$ or both hold.

Choosing $\alpha \sim \sqrt{D}$, these two quantities, $\Omega_K(\frac{D}{\alpha}) N$ and $\Omega_K(\alpha N)$, are comparable to each
other.
Choosing  $\sqrt{D} \ll \alpha \ll D$, the first quantity improves by a factor which is $\gg 1/D$ compared to the random solution, even if it is not as large as $1/\sqrt{D}$; we call this a ``pretty good solution", while the second quantity
gives an expectation value of $\sum _i(2 Z_i F_i -  Y_i \dot F_i)$ which is very large.
We will see how, in the  section which follows, to convert this large expectation value to a large expectation value (which may be positive or negative) for
$H_Z$; if this expectation value for $H_Z$ is positive then this is also a good solution, while if it is negative then it gives a solution which is
worse than random by a factor $\gg 1/\sqrt{D}$; we call this a very bad solution.  Note that $\sum_i Z_i F_i = K H_Z$.

Above, we have considered the expectation value $\langle X\rangle_T$, but we can also consider
higher moments of $X$. We will explain the reason for considering this later.
The time evolution conserves the quantity $H=X+\frac{\alpha}{D} H_Z$, but it also conserves all moments of this quantity.  Note that in the state $\pplus$ we have $\langle (H-N)^2 \rangle_+= (\alpha^2/D^2)  \langle H_Z^2 \rangle_+=(\alpha^2/D^2) N_T$.
Hence,
$\langle \tau_T^H((H-N)^2)\rangle_+=(\alpha^2/D^2) N_T$.
By Cauchy-Schwarz,
\begin{eqnarray}
\langle \tau_T^H((H-N)^2)\rangle_+&=& \langle \tau_T^H((X-N)^2) \rangle_+ +2\frac{\alpha}{D} \langle \tau_T^H((X-N) H_Z) \rangle_+
+\frac{\alpha^2}{D^2} \langle \tau_T^H(H_Z^2) \rangle_+ \\ \nonumber
&\geq &  \langle \tau_T^H((X-N)^2) \rangle_++
\frac{\alpha^2}{D^2}\langle \tau_T^H(H_Z^2) \rangle_+-
2 \frac{\alpha}{D}\sqrt{ \langle \tau_T^H(H_Z^2) \rangle_+
\langle \tau_T^H((X-N)^2) \rangle_+} \\ \nonumber
&=&\Bigl(
\sqrt{\langle \tau_T^H((X-N)^2) \rangle_+}-
\frac{\alpha}{D} \sqrt{\langle \tau_T^H(H_Z^2) \rangle_+} \Bigr)^2.
\end{eqnarray}
Hence,
\be
\label{Hvar}
 \sqrt{\langle \tau_T^H((X-N)^2) \rangle_+} \leq \frac{\alpha}{D}\sqrt{N_T} +\frac{\alpha}{D} \sqrt{\langle \tau_T^H(H_Z^2) \rangle_+}.
\ee

Hence, we have related fluctuations in $X-N$ to fluctuations in $H_Z$.
Suppose
it is the case that with probability at most $(\alpha T^2/D)^2$ that $\tau_T^H(H_Z)$ is measured to be greater than $\alpha T^2 N$ (if this does not hold, then we can repeat the algorithm polynomially many times to have a large probability of obtaining a state in the computational basis with expectation value of $H_Z$ greater than $\alpha T^2 N$).
Under this assumption, then
since $\Vert H_Z \Vert\leq DN$, it follows that
$\sqrt{\langle \tau_T^H(H_Z^2) \rangle_+}=O(\alpha T^2 N)$,
and so
\be
\sqrt{\langle \tau_T^H((X-N)^2) \rangle_+}=O\Bigl(\frac{\alpha^2 T^2}{D} N+\frac{\alpha}{D} \sqrt{N_T}\Bigr).
\ee
In the limit of large $N$, the quantity $\sqrt{N_T}$ is asymptotically only $\sqrt{N}$ and so is negligible compared to the leading term, i.e., the rms (root-mean-square) fluctuations in $X-N$ are comparable to or smaller than the magnitude of $X-N$.

\section{Combining solutions}
\label{rounding}
Here we give some general results on how, given a solution to an optimization problem for a polynomial in several vectorial variables, one can construct a solution to the same problem where all variables are chosen to be the same; we call this ``combining".  Theorem \ref{roundth} is the main result.
We will use this result in both the classical and quantum algorithms; the vectors $\vec w_a$ are the solution to the problem using several vectorial variables, while the $\vec u$ is the solution with all variables the same.  This plays a key role in the classical algorithm, while for the quantum algorithm one can use a large expectation value for a quantity like $Y_i \dot F_i$, which is a polynomial in variables $Y_i, Z_i$ to
find a solution with large expectation in a single variable.

These results involve polynomials in real variables.  However, the objective function $H_Z$ is an order-$K$ polynomial in variables $Z_i \in \{-1,+1\}$.  Each $Z_i$ is chosen from $\{-1,+1\}$.  Let $\vec Z$ be a vector of choices of variables $Z$.
We write $H_Z(\vec Z)$ to denote the value of $H_Z$ for that given set of choices.

To apply the results to $H_Z$,
we randomly round choices of $Z_i$ from the interval $[-1,+1]$ to choices of $Z_i$ from the discrete set $\{-1,+1\}$ while
preserving expectation value.  Formally, consider a vectorial variable $\vec v$ with each entry chosen from the interval $[-1,+1]$.  Then, independently choosing each $Z_i$ at random from $\{-1,+1\}$, picking the probability for each $Z_i$ so that
$\expec[Z_i]=v_i$, we have that
$\expec[H_Z(\vec Z)]=H_Z(\vec v)$.

Now, let us define a polynomial $H_Z(\vec v_1,\vec v_2,\ldots,\vec v_K)$ which depends upon $K$ different vectorial variables as follows.
This polynomial will be homogeneous of order $1$ in each variable.
For each term in $H_Z$ of the form $c Z_{i_1} Z_{i_2} \ldots Z_{i_K}$, where $c$ is a scalar and $i_1,i_2,\ldots,i_K$ are a sequence of distinct choices of $i$, we have a corresponding term in
$H_Z(\vec v_1,\ldots,\vec v_K)$ equal to
$$c \frac{1}{K!} \sum_\pi (\vec v_1)_{i_{\pi(1)}} (\vec v_2)_{i_{\pi(2)}} \ldots (\vec v_K)_{i_{\pi(K)}},$$
where the sum is over permutations $\pi$ on $K$ elements and $(\vec v_a)_b$ denotes the $b$-th entry of vector $\vec v_a$.
For example, for $K=2$, given a term $-Z_2 Z_3$ we have the corresponding term $-(1/2) (\vec v_1)_2 (\vec v_2)_3-(1/2) (\vec v_1)_3 (\vec v_2)_2$.
Here in an abuse of notation we use the same symbol $H_Z(\cdot)$ for two different functions, one depending on $K$ vectorial arguments and one depending on a single vectorial argument.

Note that
\be
H_Z(\vec v,\vec v,\ldots,\vec v)=H_Z(\vec v).
\ee
We will show that, given
choice of $\vec v_1,\ldots,\vec v_K$ such that $H_Z(\vec v_1,\ldots,\vec v_K)$ has a certain magnitude,
we will find a choice of $\vec v$ such that $H_Z(\vec v)$ obeys certain conditions on its magnitude.

This will then be used in the classical setting in the following simple way: we will pick some vector $\vec w_2$ at random and then
choose $\vec w_1$ greedily to optimize 
$H_Z(\vec w_1,\vec w_2,\vec w_2,\vec w_2,\ldots,\vec w_2)$.  Here
the variable $\vec w_1$ appears $1$ time while the variable $\vec w_2$ appears $K-1$ times.
This will give us the choice of $K$ different vectorial variables (though one variable is repeated $K-1$ times) from which we will construct a solution with a single variable.

Item 1 in the theorem will be the case that we need most.  Item 2 almost follows from item 1 with $\epsilon=1$, except item 2 has slightly tighter bounds.  Item 3 is given for completeness as it shows that some similar results hold when many variables are present and also item 3 is used in the proof of item 1.  Thus, the reader may consider only item 1.
\begin{theorem}
\label{roundth}
Let $P(\vec v_1,\vec v_2,\ldots,\vec v_K)$ be a polynomial in vectorial variables $\vec v_1,\ldots,\vec v_K$ which is homogeneous of order $1$ in each argument so that
\be
P(\vec v_1,\ldots,\vec v_K)=\sum_{i_1,\ldots,i_K} a_{i_1,\ldots,i_K} \prod_a (\vec v_a)_{i_a},
\ee
 where $(\vec v_a)_i$ denotes the $i$-th entry of vector $\vec v_a$.
 
 Assume that all vectors $\vec v_a$ have the same number of entries, and assume that $P$ is symmetric under permuting its arguments, i.e., that $a_{i_1,\ldots,i_K}$ is symmetric under permuting its arguments.
 
Then the following holds:
\begin{itemize}
\item[1.]  Suppose that there exist vectors $\vec w_1,\vec w_2$ such that
 $P(\vec w_1,\vec w_2,\vec w_2,\vec w_2,\ldots,\vec w_2)=C/K$.
 Then for any $\epsilon>0$, at least one of the following two possibilities holds: 
 \begin{itemize}
 \item[A:]
 there exists some vector $\vec u$ with $|\vec u_i|\leq |(\vec w_1)_i|+|(\vec w_2)_i|$ for all $i$
 such that
  \be
 P(\vec u,\vec u,\ldots,\vec u) \geq P(\vec w_2,\vec w_2,\ldots,\vec w_2)+\epsilon C \cdot \Omega(1)
 \ee
 or
 \item[B:]
 there exists some vector $\vec u$ with $|\vec u_i|\leq |(\vec w_1)_i|+|(\vec w_2)_i|$ for all $i$
 such that
  \be
 |P(\vec u,\vec u,\ldots,\vec u)| \geq C \cdot \exp(-O(K))/\epsilon.
  \ee
  \end{itemize}
Remark: item A is a statement about $P$ while item B is a statement about the absolute value of $P$.
 
 \item[2.] Suppose that there exist vectors $\vec w_1,\vec w_2$ such that
 $P(\vec w_1,\vec w_2,\vec w_2,\vec w_2,\ldots,\vec w_2)=C/K$
 and such that $|(\vec w_a)_i|\leq 1$ for all $a,i$.  (That is, the variable $\vec w_1$ appears $1$ time while the variable $\vec w_2$ appears $K-1$ times.  Then, there exists some vector $\vec u$ with $$|\vec u_i|\leq |(\vec w_1)_i|+|(\vec w_2)_i|$$ for all $i$ such that
 \be
 |P(\vec u,\vec u,\ldots,\vec u)| \geq P(\vec w_2,\vec w_2,\ldots,\vec w_2)+C \cdot \Omega(1/K).
 \ee
 
 \item[3.]  Suppose that there exist some vectors $\vec w_1,\ldots,\vec w_K$ such that $P(\vec w_1,\ldots,\vec w_K)=C$
and such that $|(\vec w_a)_i|\leq 1$ for all
 $a,i$.  Then, there exists some vector $\vec u$ with $$|\vec u_i|\leq 1$$ for all $i$ such that
 \be
 |P(\vec u,\vec u,\ldots,\vec u)|\geq 
 \frac{K!}{K^K}C .
 \ee
 \end{itemize}
 
 Further, in all cases, we can find $\vec u$ up to any desired nonzero error in a time linear in $N$, exponential in $K$, and at most polynomial in inverse error multiplied by the magnitude of the terms in the polynomial.
\end{theorem}

Note that item 3 above allows all of the $\vec w_a$ to be distinct.  Items 1,2 consider the case of just two different $\vec w_a$, with $\vec w_2$ repeated $K-1$ times in the argument of $P(\cdot)$.  We can summarize item 2 as saying that one can obtain a solution whose absolute value is
close to $C$, while item 1 can be summarized for small $\epsilon$ as saying that, compared to $P(\vec w_2,\vec w_2,\ldots,\vec w_2)$, either we can improve by a small amount (this is the ``pretty good") or there is a solution which is much worse (this is the ``very bad").
Note also that the bound on $|(\vec u)_i|$ is different in item 2 compared to items 1,3.

We now prove the theorem.
Define a function $\vec u(\cdot)$, from ${\mathbb R}^2$ to vectors, by
\be
\vec u(x_1,x_2)=\sum_a x_a \vec w_a,
\ee
where $x_a \vec w_a$ denotes the vector with $i$-th entry equal to $x_a (\vec w_a)_i$.

First, we prove item 1.
We need
\begin{lemma}
\label{largecoefflemma}
Let $p(x)=\sum_{0 \leq i \leq d} a_i x^i$ be an order-$d$ polynomial in real variable $x$.  Let $p_{max}={\rm max}_{x \in [-1,1]} p(x)$.
Let $a_{max}={\rm max}_{i \geq 1} |a_i|$.
Then,
\be
p_{max} \geq a_0+(1/6) a_1^2/a_{max}.
\ee

Remark: the factor $1/6$ in the above equation is not optimal.  It can be tightened easily.  Indeed, for $a_1 \ll a_{max}$, the factor $1/6$ approaches $1/2$.
\begin{proof}
Consider $p(x_0)$ for $x_0=a_1/(4a_{max})$.  We have
$p(x_0)=a_0+(1/4)a_1^2/a_{max}+\sum_{2\leq i \leq d} a_i x_i^i$.
So,
\begin{eqnarray}
p(x_0) &\geq & a_0+(1/4) a_1^2/a_{max} -\sum_{2\leq i \leq \infty} a_{max} \cdot \Bigl(\frac{|a_1|}{4a_{max}}\Bigr)^i \\ \nonumber
&\geq & a_0 + (1/4) a_1^2/a_{max}-\sum_{2\leq i \leq \infty} 
a_{max} \cdot \Bigl(\frac{|a_1|}{4a_{max}}\Bigr)^2 \cdot \Bigl(\frac{|a_1|}{4a_{max}}\Bigr)^{i-2}
\\ \nonumber
&\geq & a_0  + (1/4) a_1^2/a_{max}-a_{max} \cdot \Bigl(\frac{|a_1|}{4a_{max}}\Bigr)^2 \sum_{2 \leq i \leq \infty} (1/4)^{i-2}  \\ \nonumber
&\geq & a_0 + (1/6) a_1^2/a_{max}.
\end{eqnarray}
\end{proof}
\end{lemma}

Define polynomial $Q(x)\equiv P(\vec u(x,1),\ldots,\vec u(x,1))$.  Apply lemma \ref{largecoefflemma} to $Q(x)$ with $a_1=C$.  If
case A of item 1 of theorem \ref{roundth} does not hold for some given $\epsilon$, then
$(1/6) C^2/a_{max}<\epsilon C$ so $a_{max}>(1/6) (C/\epsilon)$.  So for some $i\geq 1$, $|a_i|>(1/6) (C/\epsilon)$.
So, $$|a_i|={K \choose i} |P(\vec w_1,\ldots,\vec w_1,\vec w_2,\ldots,\vec w_2)|>(1/6) (C/\epsilon),$$ where $\vec w_1$ appears $i$ times in the argument of $P(\cdot)$ and $\vec w_2$ appears $K-i$ times.
So, by item 3 of theorem \ref{roundth}, which we prove below, there is some choice of
 $\vec u$ with $|(\vec u)_i|\leq 1$ for all $i$ such that
$$ |P(\vec u,\vec u,\ldots,\vec u)|\geq \frac{1}{{K \choose i}}
 \frac{K!}{K^K}(1/6) (C/\epsilon) .$$
 Since $\frac{1}{{K \choose i}}
 \frac{K!}{K^K}(1/6)\geq \exp(-O(K))$, the result follows.
 
 This completes the proof.

We next prove item 2.  We need the lemma:
\begin{lemma}
\label{goodorbadlemma}
Let $p(x)$ be a polynomial of order $K$ with $p(x)=\sum_{0 \leq i \leq d} a_i x^i$.  Then, for $K$ odd
\be
{\rm max}_{x\in [-1,1]}\Bigl(|(p(x)|\Bigr) \geq |a_1|/K,
\ee
and for $K$ even
\be
{\rm max}_{x\in [-1,1]}\Bigl(|(p(x)|\Bigr) \geq |a_1|/(K-1),
\ee
\begin{proof}
The proof is similar to the proof that the Chebyshev polynomials minimize the maximum absolute value on the interval $[-1,1]$ among all polynomials with given {\it leading} coefficients, i.e., with given value of $a_K$.  In this case, we instead fix the value of $a_1$, but the proof is almost the same.

First, without loss of generality we may assume that $p(x)=-p(-x)$, as $(p(x)-p(-x))/2$ is also a polynomial of order $K$ with coefficient of the linear term also equal to $a_1$ and $|(p(x)-p(-x))/2| \leq {\rm max}(|p(x)|,|p(-x)|)$.  So, we can assume that $K$ is odd and the result for even $K$ will follow immediately from the result for odd $K$.

Also, without loss of generality we may assume that $a_1=1$.  Indeed, if $a_1=0$, then the result is trivially true, while for any nonzero $a_1$ we can instead consider $p(x)/a_1$.

Assume that the lemma is false, i.e., assume that $p(x)$ has maximum absolute value on the interval $[-1,1]$ which is strictly smaller than $1/K$.

Let $T_n(x)$ be the Chebyshev polynomials of first kind.  For odd $K$, we have that $-(-1)^K \cdot T_K(x)/K$ is an polynomial of order $K$ which has coefficient of the linear term equal to $1$.  Further, $-(-1)^K \cdot T_K(x)/K$ has a maximum absolute value on the interval $[-1,1]$
equal to $1/K$ and it attains this maximum $K+1$ times on this interval at points $x=\cos(k\pi/K)$ for $0\leq k \leq K$.  Let
$q(x)=p(x)+(-1)^K \cdot T_K(x)/K$.  So, $q(x)$ has coefficient of the linear term equal to zero, i.e., since it is an odd function of $x$, we have
$q(x)=\sum_{i=3,5,\ldots,K} b_i x^i$ for some coefficients $b_i$.  Further by the assumption that $p(x)$ has absolute value strictly smaller than $1/K$ on the interval, we have that at points $x=\cos(k\pi/K)$ the sign of $q(x)$ is the same as the sign of $(-1)^K \cdot T_K(x)/K$.  So, since the sign of $T_K(x)$ alternates at these points, i.e., the sign for even $k$ is opposite to that for odd $k$, we have that $q(x)$ changes sign at least $K$ times so $q(x)$ must have at least $K-1$ distinct zeros.  However, $q(x)$ has order $K$ and the root at $x=0$ is triply degenerate so in fact $q(x)$ can only have at most $K-2$ distinct zeros, giving a contradiction.
\end{proof}
\end{lemma}

Define polynomial $Q(x)\equiv P(\vec u(x,1),\ldots,\vec u(x,1))$.  Applying lemma \ref{goodorbadlemma} to $p(x)=Q(x)$ with $a_1=C$, the result follows.

For both item 1 and 2,
we can find an $x$ which maximizes or minimizes $|Q(x)|$ up to any given error by exhaustively trying a discrete set of points on the interval $[-1,1]$ with the spacing between points dependent on the error.

We finally prove item 3 of theorem \ref{roundth}.  We need:
\begin{lemma}
\label{coeffroundlemma}
Let $p(x_1,\ldots,x_K)$ be a polynomial (not necessarily homogenous) of order at most $K$ in real variables $x_1,\ldots,x_K$.
Suppose that the coefficient of the term $\prod_i x_i$ in $p(\cdot)$ is equal to $C$.
Then, for some choice of $x_1,\ldots,x_K \in \{-1,+1\}^K$ we have that $|p(x_1,\ldots,x_K)|\geq C$.
\begin{proof}
We claim that
\be
C=\frac{1}{2^K} \sum_{x_1,\ldots,x_K \in \{-1,+1\}^K} \Bigl(\prod_i x_i \Bigr) \cdot p(x_1,\ldots,x_K).
\ee
This holds because any term in $p(\cdot)$ proportional to $\prod_i x_i^{d_i}$ for some sequence of integers $d_i$ will vanish in the weighted sum above unless all $d_i$ are odd.  However, since $p(\cdot)$ has order $d$, the only such nonvanishing term is that with all $d_i=1$.

Hence, $|C|\leq {\rm max}_{x_1,\ldots,x_K \in \{-1,+1\}^K}(|p(x_1,\ldots,x_K)|)$.
\end{proof}
\end{lemma}

To prove item 3, in an abuse of notation let us redefine $\vec u(\cdot)$ to 
denote a function from ${\mathbb R}^K$ to vectors, by
\be
\vec u(x_1,\ldots,x_K)=\sum_a x_a \vec w_a.
\ee
Consider polynomial $Q(x_1,\ldots,x_K)\equiv P(\vec u(x_1,\ldots,x_K),\ldots,\vec u(x_1,\ldots,x_K))$.
The polynomial $Q(\cdot)$ is of order $K$ and the coefficient of
$\prod_i x_i$ in $Q(\cdot)$ is equal to $C K!$.
So, by lemma \ref{coeffroundlemma}, there exists some choice of $x_1,\ldots,x_K\in \{-1,1\}^K$ such that
$|Q(x_1,\ldots,x_K)| \geq C K!$.
Set $\vec u=\frac{1}{K} \vec(x_1,\ldots,x_K)$ so that indeed $|(\vec u)_i)|\leq 1$ for all $i$.

Then, $|P(\vec u,\ldots,\vec u)|\geq (1/K)^K C K!.$
This prove item 3 and trivially one can find the choice of $\vec u$ by iterating over the $2^K$ possible choices of $x_1,\ldots,x_K \in \{-1,1\}^K$.

\section{Classical Algorithm}
\label{ca}
We now describe the classical optimization algorithm.
Recall that we define $F_i$ to equal $Z_i$ times the sum of terms in $H_Z$ that include $Z_i$. 

\begin{algorithm}
\caption{Classical algorithm}
\begin{itemize}
\item[{\bf 1.}] 
Choose a set $S$ of bits, by including each bit in $S$ independently with probability $1/2$.

\item[{\bf 2.}] Define vectorial variables $\vec w_1,\vec w_2$ as follows; the index of the vectorial variable will range over $\{1,\ldots,N\}$ so that it labels a bit.  Let $\vec w_2$ be a vector with $(\vec w_2)_i=0$ for $i \in S$ while for $i \not \in S$ we choose $(\vec w_2)_i$ to be $+1$ or $-1$ independently and uniformly at random.
We choose vector $\vec w_1$ so that $(\vec w_1)_i=0$ for $i\not\in S$ while for $i\in S$ we choose $(\vec w_1)_i$ ``greedily".
That is, we pick $(\vec w_1)_i=+1$ if $F_i(\vec w_2)>0$ and $(\vec w_2)_i=-1$ otherwise.

\item[{\bf 3.}] Finally, apply item 1 of theorem \ref{roundth}.
By this item, for any $\epsilon>0$, we can either find a choice of $\vec u$ such that
$H_Z(\vec u)\geq H_Z(\vec w_2) + \epsilon C \cdot \Omega(1)$
 or
 such that
 $|H_Z(\vec u)|\geq  C \cdot \exp(-O(K))/\epsilon$,
 where $C=\sum_{i \in S} |F_i|$.
 \end{itemize}
\label{classical}
\end{algorithm}

\subsection{Some Probability Bounds}
We collect here some probability bounds that we will need to analyze this algorithm, as well as to analyze the classical algorithm.
The use of these bounds is similar to that in Ref.~\cite{barak2015beating}.

By theorem 9.23 of Ref.~\cite{o2014analysis}, for any function $f$ of order at most $K$ from $\{-1,1\}^N\rightarrow \mathbb{R}$ we have for any
$t \geq (2e)^{K/2}$ that
\be
\label{boundeq}
\pr_{x \in \{-1,1\}^N}[|f(x)|\geq t \expec[|f|^2]^{1/2}] \leq \exp(-\frac{K}{2e} t^{2/K}).
\ee

By theorem 9.24 of Ref.~\cite{o2014analysis}, for any nonconstant function $f$ of order at most $K$ from $\{-1,1\}^N\rightarrow \mathbb{R}$ ,
\be
\label{noncon0}
\pr_{x \in \{-1,1\}^N}[f(x) >\expec[|f|]] \geq \frac{1}{4}\exp(-2K).
\ee

Hence,  for any nonconstant function $f$ of order at most $K$ from $\{-1,1\}^N\rightarrow \mathbb{R}$ , by applying Eq.~(\ref{noncon0}) to $f^2$ we have
\be
\label{noncon}
\pr_{x \in \{-1,1\}^N}[|f(x)| >\expec[|f|^2]^{1/2}] \geq \frac{1}{4}\exp(-4K).
\ee

\subsection{Analysis of Classical Algorithm}
We will use $\expec[\ldots]$ to denote expectation values over choices of $\vec w_2$.
We claim that $\expec[|F_i|]\geq \sqrt{D}\exp(-O(K))$ and that $\expec[C]\geq N\sqrt{D} \exp(-O(K))$.  To see this, note that each site is in $S$ with probability at least $1/2$.  For any site (including a site in $S$ in particular), 
we have that $\expec[F_i(\vec w_2)^2]$ is equal to $2^{-(K-1)} D$.  The factor of $2^{-(K-1)}$ arises because each monomial in $F_i$ is of order $K-1$ and has probability $2^{-(K-1)}$ that all bits in that monomial are not in $S$.
So, $\expec[|F_i|]\geq \sqrt{D}\exp(-O(K))$  follows from Eq.~(\ref{noncon}).

Note that the maximal value for $C$ is $ND$, so with probability at least $1/{\rm poly}(D)$ we find a choice of $\vec w_2$ such that
$C$ is at least a constant factor times the expected value.  Here the constant factor can be any number strictly less than $1$, for example $1/2$.
This is an application of Markov's inequality.  Consider $ND-C$ as a non-negative random variable with expectation value $ND-\expec[C]$.  The probability that $C$ is smaller than $\expec[C]/2$, for example, is bounded by $(ND-\expec[C])/(ND-\expec[C]/2)=1-\Omega(1/\sqrt{D}) \exp(-O(K))$ so with probability $\Omega(1/\sqrt{D})\exp(-O(K))$ we have that $C \geq \expec[C]/2$.

For such choices of $\vec w_2$,
the algorithm must choose either case 1A or case 1B at least half the time (or any other number $\Omega(1)$ rather than one half).
Hence, at least one of the following holds: with probability $P$ at least ${\rm poly}(1/D)$ the algorithm chooses case 1A  and $C$ is within a constant factor of the expected value so that
$H_Z(\vec u)$ is
at least $H_Z(\vec w_2)+N\epsilon \sqrt{D}\exp(-O(K))$, or,  with probability $P$ at least ${\rm poly}(1/D)$
the 
algorithm chooses case 1B 
and $C$ is within a constant factor of the expected value so that
 $|H_Z(\vec u)| \geq N\sqrt{D}\exp(-O(K))/\epsilon$.

Now consider $H_Z(\vec w_2)$.  This has expectation value $0$ and the expectation value of $H_Z(\vec w_2)^2$ is $O(N_T)$.  So by Eq.~(\ref{boundeq})
the probability that $|H_Z(\vec w_2)|$ is larger than $O(\log(N)^{K/2} \sqrt{N_T})$ is equal to  
$N^{-K/2e}$.
This probability $N^{-K/2e}$ is asymptotically (in $N$) negligible compared to $P$ for any $P=\Omega({\rm poly}(1/D))$.
Hence, by a union bound, with probability $P=\Omega({\rm poly}(1/D))$ one of the cases in the above paragraph holds (i.e., algorithm chooses either case 1A or 1B and the given bounds on $H_Z(\vec u)$ hold) and also $|H_Z(\vec w_2)|$ is $o(N)$.
So,
\begin{theorem}
\label{dualclemma}
At least one of the following holds:
with probability ${\rm poly}(1/D)$,
the algorithm either chooses case 1A with
$H_Z(\vec u)$
at least 
$N\epsilon \sqrt{D}\exp(-O(K))-
O(\log(N)^{K/2} \sqrt{N_T})=
N\epsilon \sqrt{D}\exp(-O(K))-o(N)$, or,
with probability $P$ at least ${\rm poly}(1/D)$
the 
algorithm chooses case 1B 
with
$|H_Z(\vec u)| \geq N\sqrt{D}\exp(-O(K))/\epsilon$.

In the second case, either $H_Z(\vec u) \geq N\sqrt{D}\exp(-O(K))/\epsilon$ or
$H_Z(\vec u) \leq -N\sqrt{D}\exp(-O(K))/\epsilon$ so
that for any $\epsilon =O(1)$, either the algorithm
finds a solution with 
$H_Z(\vec u)\geq
N\epsilon \sqrt{D}\exp(-O(K))-o(N)$ or it finds a solution
with $H_Z(\vec u)\leq -N \sqrt{D} \exp(-O(K))/\epsilon$.

Here we adopt a convention that for quantities which are negligible for large $N$, such at $O(\log(N))^{K/2} \sqrt{N_T}$, we simply write $o(N)$, rather than giving the detailed dependence on $K,D$.
\end{theorem}

Note that for odd $K$, we can guarantee that it achieves expected $H_Z(\vec u)\geq N\sqrt{D} \exp(-O(K))$ as in Ref.~\cite{barak2015beating} since we can pick $\epsilon=1$ and if case 1B occurs, we can change the sign of all bits.

\subsection{Modification With Generalized Duality and Comparison to Quantum Duality}
The classical algorithm above achieves  a duality very similar to that
of the conjectured duality of conjecture \ref{conjer} with $\alpha^2 T^2/D=1$.
Set $$\epsilon \sqrt{D}=\alpha T^2.$$  Then,
$$\alpha=\frac{\alpha^2 T^2}{D} \cdot \frac{D}{\alpha T^2}=1 \cdot \frac{\sqrt{D}}{\epsilon}=\frac{\sqrt{D}}{\epsilon}.$$
Thus, while we conjectured that the quantum algorithm
either gave an expectation value of $H_Z$ equal to $\Omega_K(1) \alpha T^2 N$
or that there was some state with $\sum _i (2Z_i F_i - Y_i \dot F_i)$ at least $\Omega_K(1) \alpha N$, we find for the classical algorithm that
we can prove either that at least ${\rm poly}(1/D)$ of the time it chose 1A and has
an expectation value of $H_Z$ equal to $\Omega_K(1) \alpha T^2 N$,
or that there was some state with expectation value of $H_Z$ at least $\Omega_K(1) \alpha N$ in absolute value.

The main difference then between the conjectured result for the quantum algorithm and the proven result for the classical algorithm (in the particular case that $\alpha^2 T^2/D=1$) is the replacement of $\sum_i (2 Z_i F_i - Y_i \dot F_i)=2KH_Z-\sum_i Y_i \dot F_i$ with $H_Z$ in the expectation value in the classical case.
These two operators, $\sum Y_i \dot F_i$ and $H_Z$, are closely related to each other, with the first operator $\sum_i Y_i \dot F_i$ being obtained by taking each monomial in the sum defining $H_Z$ and replacing two of the Pauli $Z$ operators in that monomial with Pauli $Y$ operators; each term in $H_Z$ then gets replaced with $\approx K^2/2$ different terms.  We describe this in more detail in subsubsection \ref{ss1}.

One may then ask if one can achieve a similar duality in the classical algorithm that would be analogous to the case of $\alpha^2 T^2/D \ll 1$.
Indeed, this can be done, as we describe in subsubsection \ref{ss2}.  We emphasize that that subsubsection considers a statement about the classical algorithm: it proves that either the {\it classical} algorithm attains a certain performance or a {\it quantum} state (not necessarily at all related to the quantum state considered in the quantum algorithm) has a certain expectation value of $H_Z$ and {\it also} that state has a certain expectation value of $X$.  The point of that subsubsection is to show that an apparent extra feature of the duality in the quantum case (when one considers $\alpha^2 T^2/D \ll 1$ so that the expectation value of $X$ is large) does not actually directly give any more powerful results.

\subsubsection{Combining Solutions for Quantum Algorithm}
\label{ss1}
Suppose we have a quantum state with large (in absolute value) expectation value for the operator $\sum_i (2 Z_i F_i - Y_i \dot F_i)$.  We will describe how to construct a classical state which has large expectation value (again, in absolute value) for $H_Z$.  
Indeed, the classical state will be constructed simply by measuring the quantum state in a product basis and (possibly) combining solutions.

Let $V$ denote the absolute value of the expectation value of $\sum_i (2 Z_i F_i - Y_i \dot F_i)$ in the quantum state.  Then, at least one of the following holds: the expectation value of $2\sum_i Z_i F_i$ is at least $V/2$ in absolute value, or the expectation value of $-\sum_i Y_i \dot F_i$ is at least $V/2$ in absolute value.  In the first case, using the identity that $\sum_i Z_i F_i=KH_Z$, the expectation value of $H_Z$ in the state is at least $V/(4K)$ in absolute value and one can simply measure the quantum state in the $Z$ basis to obtain a classical state with expectation of $H_Z$ which is at least $V/(4K)$ in absolute value.

In the second case we have that the expectation value of $\sum_i Y_i \dot F_i$ is larger in absolute value than $V/2$.  
This operator $\sum_i Y_i \dot F_i$ is related to $H_Z$ as mentioned.  We will explore this relation in more depth.

We randomly divide the qubits into two subsets; each qubit will be placed in subset $S_1$ with probability $1/K$ and in subset $S_2$ with probability $1-1/K$, choosing independently for each qubit.  Then, we measure each qubit in $S_1$ in the $Y$ basis and measure each qubit in $S_2$ in the $Z$ basis.

We will define $\chi$ to be the state after measurement, i.e., $\chi$ is a product state where each qubit in $S_1$ is either $\pm 1$ in the $Y$ basis and each qubit in $S_2$ is either $\pm 1$ in the $Z$ basis.
Consider the expectation value of $\sum_i Y_i \dot F_i$ in this product state $\chi$.  For each term in $\sum_i Y_i \dot F_i$, the expectation value is zero unless both occurrences of Pauli $Y$ operators are for qubits in $S_1$ and all occurrences of Pauli $Z$ operators are for qubits in $S_2$; this occurs with probability $(1/K)^2 (1-1/K)^{K-2}=\Omega(1/K^2)$.
Let us use $\expec_{{\rm meas},S_1}[\ldots]$ to denote an expectation value over measurement outcomes and over choices of $S_1$.
So $\Bigl|\expec_{{\rm meas},S_1}[\langle \chi | \sum_i Y_i \dot F_i| \chi \rangle]\Bigr| \geq  \Omega(1/K^2) V$.
So, using a  similar Markov inequality as before, with at most polynomially small probability a given choice of $\chi$ and $S_1$ will have 
$|\langle \chi | \sum_i Y_i \dot F_i| \chi \rangle|$ at least $\Omega(1/K^2) V$.

Consider such a choice of $\chi$ and $S_1$.  Construct two vectors $v_1,v_2$, with $(v_1)_i = \langle \chi | Y_i | \chi \rangle$
and $(v_2)_i=\langle \chi | Z_i | \chi \rangle$.  Hence, $(v_1)_i=0$ if $i\in S_2$ and similarly $(v_2)_i=0$ if $i\in S_1$.
We will relate $\expec_{{\rm meas}} \langle \chi | \sum_i Y_i \dot F_i| \chi \rangle$ to $H_Z(\vec v_1,\vec v_1,\vec v_2,\ldots,\vec v_2)$, where $\vec v_2$ is repeated $K-2$ times.
Consider a given term $c Z_{i_1} Z_{i_2} \ldots Z_{i_K}$ in the Hamiltonian for some scalar $c$.
The corresponding term in $H_Z(\vec v_1,\vec v_1,\vec v_2,\ldots,\vec v_2)$ is
$c \frac{1}{K!} \sum_\pi (\vec v_1)_{i_{\pi(1)}} (\vec v_1)_{i_{\pi(2)}} (\vec v_2)_{i_{\pi(3)}} \ldots (\vec v_2)_{i_{\pi(K)}}$.
This vanishes unless exactly two elements of the sequence $i_1,\ldots,i_K$ are in $S_1$ and the remaining elements are in $S_2$.  Let us permute the order of the sequence so that $i_1,i_2\in S_1$.
Then, the term is equal to (after summing over permutations) $c {K \choose 2}^{-1} (\vec v_1)_{i_1} (\vec v_1)_{i_2} (\vec v_2)_{i_3} \ldots (\vec v_2)_{i_K}$.
Now consider the term in $\sum_i Y_i \dot F_i$ corresponding to 
$c Z_{i_1} Z_{i_2} \ldots Z_{i_K}$.  This is equal to a sum of ${K \choose 2}$ different terms, corresponding to the different ways of replacing two Pauli $Z$ operators with Pauli $Y$ operators.
The only one of these terms which is nonvanishing in the expectation value $\langle \chi | \sum_i Y_i \dot F_i| \chi \rangle$ is when we choose to replace $i_1,i_2$ by Pauli $Y$ operators.
In this case, the contribution to the expectation value is $2  (\vec v_1)_{i_1} (\vec v_1)_{i_2} (\vec v_2)_{i_3} \ldots (\vec v_2)_{i_K}$.
Hence, summing over all terms in $H_Z$, we find that
$H_Z(\vec v_1,\vec v_1,\vec v_2,\ldots,\vec v_K)=\frac{1}{K(K-1)}  \langle \chi | \sum_i Y_i \dot F_i| \chi \rangle$.
At this point, if desired, one could apply the combining solution techniques to $\vec v_1,\vec v_2$ to obtain a single vector with large expectation value for $H_Z$ in absolute value.

\subsubsection{Generalized Duality for Classical Algorithm}
\label{ss2}
In this subsubsubsection, we describe
a modification of the classical algorithm which provably achieves a duality similar to that in the quantum case, so that the performance of the classical algorithm is guaranteed unless there exists a quantum state with certain properties, including a large expectation value of $X$.
The modification is simple: we change step 2 to give the following modification.  The bounds in step 3 change as a consequence and are given later.

\begin{algorithm}
\caption{Modified classical algorithm}
\begin{itemize}
\item[{\bf 1.}] Fix some real number $0<p<1$.
Choose a set $S$ of bits, by including each bit in $S$ independently with probability $1/2$.

\item[{\bf 2.}] Define vectorial variables $\vec w_1,\vec w_2$ as follows; the index of the vectorial variable will range over $\{1,\ldots,N\}$ so that it labels a bit.  Let $\vec w_2$ be a vector with $(\vec w_2)_i=0$ for $i \in S$ while for $i \not \in S$ we choose $(\vec w_2)_i$ to be $+1$ or $-1$ independently and uniformly at random.
We choose vector $\vec w_1$ so that $(\vec w_1)_i=0$ for $i\not\in S$ while for $i\in S$ we choose $(\vec w_1)_i$ 
as follows.  Pick
$$(\vec w_1)_i=p \frac{F_i}{\sqrt{D}},$$
where the constant $p>0$ is chosen below.
If this choice of $(\vec w_1)_i$ gives $|(\vec w_1)_i|>1$, then replace $(\vec w_1)_i$ with $(\vec w_1)_i/|(\vec w_1)_i|$.

\item[{\bf 3.}] Finally, apply item 1 of theorem \ref{roundth}.

 \end{itemize}
\label{classicalm}
\end{algorithm}

We will always pick $$p\leq (2ec)^{-K/2},$$ where the constant $c$ is chosen below.
For any choice of $S$, the expectation value of $F_i^2$ is bounded by $D$.
So, the probability that $|p \frac{F_i}{\sqrt{D}}|>1$ is bounded by the probability that
$|F_i|>(2ec)^{K/2}\sqrt{D}$ and so by Eq.~(\ref{boundeq}), this probability is bounded by
$\exp(-cK)$.

First let us estimate $\expec[\vec F \cdot \vec w_1]$, where $\vec F$ is a vector with components $\vec F_i$.  
This is at least equal to the sum over sites $i\in S$ of
with $|(\vec w_1)_i|<1$ of the expected value of
$\expec[F_i (\vec w_1)_i]$,
which in turn is equal to the sum over sites $i\in S$ of
$$\int_{-\sqrt{D}/p}^{+\sqrt{D}/p} p\frac{f^2}{\sqrt{D}} {\rm d}\mu_i(f),$$
where $\mu_i(f)$ is the probability distribution of force $f$ on site $i$.
Using Eq.~(\ref{boundeq}), we have that $\int_{|f|>\sqrt{D}/p}{\rm d}\mu_i(f)\leq \exp(-cK)$, as explained above.
Indeed, further application of Eq.~(\ref{boundeq}) shows that  $$\int_{|f|>\sqrt{D}/p}f^2 {\rm d} \mu_i(f) \leq \exp(-cK) {\rm poly}(c).$$  To show this one may, for example, divide the integral of $|f|>\sqrt{D/p}$ into integrals over $|f|$ in intervals $[k\sqrt{D}/p,(k+1)\sqrt{D/p}]$ for integer $k$, and separately bound each integral by $(k+1)^2 (D/p) \int_{|f|>k\sqrt{D}/p} {\rm d}\mu_i(f)$.  For large enough $c$ we then have
$$\int_{-\sqrt{D}/p}^{+\sqrt{D}/p} f^2 {\rm d}\mu_i(f) \geq D \exp(-O(K)).$$
Hence,
\be
\expec[\sum_i F_i (\vec w_1)_i]\geq p \sqrt{D} \exp(-O(K)).
\ee

Hence, the constant $C$ has
$\expec[C]\geq p N \sqrt{D} \exp(-O(K)).$
As before, the maximal value for $C$ is $ND$, so with probability at least $1/{\rm poly}(D)$ we find a choice of $\vec w_2$ such that
$C$ is within a constant factor of the expected value.

For such choices of $\vec w_2$,
the algorithm must choose either case 1A or case 1B at least half the time (or any other number $\Omega(1)$ rather than one half).
Hence, at least one of the following holds: with probability $P$ at least ${\rm poly}(1/D)$ the algorithm chooses case 1A  and $C$ is within a constant factor of the expected value so that
$H_Z(\vec u)$ is
at least $o(N)+Np\epsilon \sqrt{D}\exp(-O(K))$, or,  with probability $P$ at least ${\rm poly}(1/D)$
the 
algorithm chooses case 1B 
and $C$ is within a constant factor of the expected value so that
the $|H_Z(\vec u)| \geq
o(N)+Np\sqrt{D}\exp(-O(K))/\epsilon$.

Thus far, it seems that all we have accomplished is worsening the previous result (by a factor of $p$).  However, now we show how if the second case holds (the algorithm chooses case 1B), then we can construct a {\it quantum} state with large expectation value of $X$ and with an expectation value (in that quantum state) of
$|H_Z(\vec u)|$ which is 
at least $O(\log(N)^{K/2} \sqrt{N_T}) +Np\sqrt{D}\exp(-O(K))/\epsilon$.

Before defining $\psi$, let us note the following.  When the algorithm chooses case 1B, the vector $\vec u$ is equal to a linear combination $xp \vec w_1+\vec w_2$ for some $x\in [-1,1]$.  Discretizing the interval $[-1,1]$ into ${\rm poly}(D)$ bins, each of width ${\rm poly}(1/D)$, we find
that if the algorithm chooses case 1B with probability at least ${\rm poly}(1/D)$, then there is some bin such that with probability at least ${\rm poly}(1/D)$ the algorithm chooses case 1B and such that $x$ falls in that bin.  Choosing the width of the bins small enough, we can assume then
that there is some {\it fixed} value of $x=x_0\in [-1,1]$ such that with probability ${\rm poly}(1/D)$ the vector
$\vec u=x_0p \vec w_1 + \vec w_2$ has
$|H_Z(\vec u)| \geq o(N)+Np\sqrt{D}\exp(-O(K))/\epsilon-N/{\rm poly}(D)$.

Now apply Eq.~(\ref{boundeq}) to this case.  Let $\expec[H_Z]_{x_0}$ denote the expectation value of $H_Z (\vec u)$ for vector $\vec u=x_0 \vec w_1 + \vec w_2$, taking the expectation value over choices of $\vec w_2$.
This expectation value is the expectation value of a polynomial of order at most $O(K^2)$, as each monomial in $H_Z$ is of order $K$ and each entry of $\vec w_1$ in turn is a monomial of order at most $K-1$; it is true that we cutoff entries $\vec w_1$ by $1$, but this occurs with negligible probability.
The expectation value $\sqrt{\expec[H_Z^2]_{x_0}}$ is $O(N {\rm poly}(D))$. 
Hence, using Eq.~(\ref{boundeq}), we can bound fluctuations of $H_Z(\vec x_0p \vec w_1+\vec w_2)$
about its average.  So, since with probability at least 
${\rm poly}(1/D)$ we have
$|H_Z(\vec x_0p \vec w_1+\vec w_2)| \geq
o(N) +Np\sqrt{D}\exp(-O(K))/\epsilon-N/{\rm poly}(D)$,
we have that
$|\expec[H_Z]| \geq o(N) +Np\sqrt{D}\exp(-O(K))/\epsilon-N/{\rm poly}(D)$.

We now define this state $\psi$ by
\be
\psi\equiv 
\prod_{i \in S} \exp(i \theta Y_i F_{\overline S,i})
\pplus,
\ee
where $$\theta=p\frac{x_0}{\sqrt{D}}$$ and where we define $F_{\overline S,i}$ to denote the sum of all terms in $F_i$ which are supported on the complement of $S$.

We first compute $\langle \psi| H_Z | \psi \rangle$.  Consider any term in $H_Z$.  Such a term is proportional to some monomial $M\equiv Z_{i_1} Z_{i_2} \ldots Z_{i_K}$, for some sequence of distinct qubits $i_1,\ldots,i_K$.  Suppose that $i_1,\ldots,i_j$ are in $S$ and $i_{j+1},\ldots,i_K$ are in the complement of $S$.  We have
that $\langle \psi | M | \psi \rangle$ is equal to
$$\langle \pplus |
\Bigl(\prod_{l=1}^j (\sin(\theta F_{\overline S,i_l}) X_{i_l} + \cos(\theta F_{\overline S,i_l}) Z_{i_l})
\Bigr)
 \Bigl(\prod_{l=j+1}^K Z_{i_l} \Bigr)
|\pplus \rangle.$$
We can expand the above expectation value as a sum of $2^j$ different expectation values, by choosing for each $j=1,\ldots,l$ to take either
$\sin(\theta F_{\overline S,i_l}) X_{i_l}$ or $\cos(\theta F_{\overline S,i_l}) Z_{i_l}$.  However, every expectation value for which we
choose $\cos(\theta F_{\overline S,i_l}) Z_{i_l}$ for at least one choice of $l$ is vanishing in the state $\pplus$, as then $Z_{i_l}$ appears exactly once in the product (the terms $F_{\overline S,i_m}$ do not contains $Z_{i_l}$.
Hence, 
\be
\langle \psi | M | \psi \rangle=
\langle \pplus |
\Bigl(\prod_{l=1}^j \sin(\theta F_{\overline S,i_l}) X_{i_l} 
)
\Bigr)
 \Bigl(\prod_{l=j+1}^K Z_{i_l} \Bigr)
|\pplus \rangle.
\ee

We now estimate the error in approximating  $\sin(\theta F_{\overline S,i_l})$ by
$\theta F_{\overline S,i_l}$.  This error, for any $M$, is $O(\theta^3 F_{\overline S,i_l}^3)$.
We show below that this is negligible for sufficiently small $\theta$.

Before bounding this error, note that if we
include only the linear term $\theta F_{\overline S,i_l}$ in the approximation to the sine, then for $\theta=px_0/\sqrt{D}$ we find that
 the expectation value of $H_Z$ in state $\psi$ is equal to $H_Z(\vec w_1+\vec w_2)$ with $(\vec w_1)_i=px_0F_i/\sqrt{D}$ for $i\in S$.

To get oriented, suppose that $F_{\overline S,i_l}$ were bounded by $\sqrt{D}$.
Then, choosing $\theta$ sufficiently small compared to $D^{-3/4}$, and choosing $p=\theta\sqrt{D}$ we find that $\theta^3 F_{\overline S,i_l}^3$ would be
bounded by $D^{-9/4} D^{3/2}=D^{-3/4}$.  Summing over terms (there are $O(ND)$ terms) we find an error
at most $ND^{1/4}$ which is comparable to $Np \sqrt{D}$.
Now, $F_{\overline S,i_l}$ is not bounded by $\sqrt{D}$, but Eq.~(\ref{boundeq}) gives exponential decay bounds and so
for
$\theta=O(D^{-3/4}) \exp(-O(K))/{\rm polylog}(D)$,
where the logarithm depends on $K$,
we have that the error is negligible compared to $Np\sqrt{D}$.
This choice of $\theta$ requires
\be
\label{prange}
p=O(D^{-1/4})\exp(-O(K))/{\rm polylog}(D).
\ee

Hence, $|\langle \psi | H_Z| \psi \rangle|\geq o(N) +Np\sqrt{D}\exp(-O(K))/\epsilon-N/{\rm poly}(D)$.

We now consider the expectation value
$\langle \psi | X| \psi\rangle$, as well as higher moments in $X$.
Indeed, we have
$\langle \psi | X_i | \psi \rangle=1-\frac{p^2 x_0^2}{2}+\ldots$ where the $\ldots$ denote higher order terms in $\theta F_{\overline S,i_l}$.
Using the same fluctuation bounds as above, for the given range of $\theta$, these higher order terms are negligible.

Hence,
$\langle \psi | X| \psi\rangle=N-\frac{p^2 x_0^2}{2} N \Omega(1)$.
Comparing to the conjectured performance of the quantum algorithm, we see that
for
$$p \sim \frac{\alpha T}{\sqrt{D}},$$
and
$$\epsilon \sim T,$$
the two results match.
Note, however, that we only can show this result for sufficiently small $p$ obeying Eq.~(\ref{prange}) for the classical algorithm.  Of course, we can always achieve the performance of theorem \ref{dualclemma}; the restriction on $p$ here is just if we also wish to show the existence of a quantum state obeying with large expectation value of $X$.

One can consider higher moments too.  Note that $\langle \psi | X_i X_j |\psi \rangle-\langle \psi |X_i | \psi \rangle \cdot \langle \psi | X_j | \rangle$ is vanishing unless $i,j$ both appear in some term in $H_Z$ or unless there is some $k$ such that $i,k$ are both in some term in $H_Z$ and $k,j$ are both in some term in $H_Z$.  Hence, $\Bigl(\langle \psi | X^2 |\psi \rangle - \langle \psi | X | \psi \rangle^2\Bigr)=o(N)$.
Similar bounds can be made for higher moments.

\section{Large $X$ Expectation Value in Duality}
\label{mf}
We now consider some applications of these dualities.

\subsection{Random Models}
First consider a {\it random} model.  Consider any $K$ and any $D$.  We consider a fixed set of terms in $H_Z$, but with the signs of each term chosen randomly.  We choose the signs independently, setting them equal to $+1$ with probability $1/2$ and $-1$ with probability $1/2$.  Then, for any choice of $\vec v \in \{-1,+1\}^N$, the expectation value of $H_Z(\vec v)$ is equal to $0$.  The probability that $|H_Z(\vec v)|$ is greater than $\Delta$ is bounded by $\exp(-\Omega(\Delta^2/N_T))=\exp(-\Omega(\Delta^2/(ND)))$.  There are $2^N$ possible choices of $\vec v$, so by a union bound, with high probability there is no choice of $\vec v$ such that $|H_Z(\vec v)|$ is greater than $O(N\sqrt{D})$.

Hence, by theorem \ref{dualclemma} with $\epsilon=\exp(-O(K))$, we have that with high probability, a random instance has the property that the classical algorithm succeeds in finding a solution with $H_Z \geq N\sqrt{D} \exp(-O(K))$ a fraction at least ${\rm poly}(1/D)$ of the time.

\subsection{Mean-Field Treatment}
Now we consider some heuristic motivation why it may be worth considering the dualities that involve a large expectation value of $X$.

For motivation, to explain why this large expectation value of $X$ may be useful, we give an approximate mean-field treatment: consider some Hamiltonian of order $K$ that we will call $H_{0}$ that is diagonal in the $Z$-basis.
Suppose we wish to minimize the expectation value of $H_Z$ over states with given expectation value of $X$, i.e., we seek a state with large negative expectation value of $H_Z$.
If no constraint were placed on the expectation value of $X$, then we maximize $H_Z$ by choosing some state in the computational basis.
For each qubit $i$, this state has some expectation value $\langle Z_i \rangle = z_i$ with $z_i\in \{-1,+1\}$.
If we wish to obtain a nonzero expectation value of $X$, then a simple way is to take a product state, where each qubit has
$\langle X_i \rangle=\cos(\theta)$ and $\langle Z_i \rangle=z_i \sin(\theta)$, for some angle $\theta$.  For $\theta=\pi/2$, we recover the classical state.  At small $\theta$, the expectation value of $H_Z$ is proportional to $\theta^K$, while the expectation value
of $1-X_i$ is proportional to $\theta^2$.  Thus, for $K>2$, the expectation value of $H_0$ drops more rapidly as a function of $\theta$ than does the expectation value of $1-X_i$.

A similar mean-field treatment might be applied to a Hamiltonian that includes both $Y$ and $Z$ operators, such as $\sum_i Y_i \dot F_i$ relevant to the quantum algorithm: given any product solution of $H_0$ with $\langle Z_i\rangle=z_i$ and $\langle Y_i \rangle=y_i$ with $z_i^2+y_i^2=1$, we can define a product state with
$\langle X_i \rangle=\cos(\theta)$ and $\langle Z_i \rangle=z_i \sin(\theta)$ and $\langle Y_i \rangle=y_i \sin(\theta)$.

If this mean-field procedure were the best possible then we would have very strong bounds on the existence of such a state: we would have (for small $\theta$) the scaling
$\theta^2 \sim p^2$ and while the expectation value of $H_Z$ would be at most $\theta^K$ in absolute value times the minimal value of $H_Z$.
Call this optimal value $H_Z^{min}$.
So, we would have
$|H_Z^{min}| \theta^K \sim Np\sqrt{D}\exp(-O(K))/\epsilon$
while $p^2 \sim \theta^2$.
Here we are ignoring terms which are $o(N)$.

Ignoring $K$-dependent constants such as $\exp(-O(K))$, and
taking, at the most optimistic situation, $p\sim \theta\sim 1/\sqrt{D}$ (since for smaller $\theta$ the expectation value of $X_i$ is within $1/D$ of $1$ and certainly the mean-field is not accurate here),
we would find that such a state has
$|H_Z^{min}| \sim D^{K/2} N/\epsilon$.
That is, either the algorithm finds a solution with expectation value of $H_Z$ at least $N\epsilon$ or $|H_Z^{min}| \gtrsim D^{K/2} N/\epsilon$.
For the case $K=2$, this is the same guarantee as before, if we rescale $\epsilon\rightarrow \epsilon\sqrt{D}$, but for $K=4$ or larger, this is a much stronger guarantee.

Of course, this mean-field procedure is only an approximation and other states may exist with more negative expectation value of $H_Z$ at the same expectation value of $X$.

Still, one use of the large expectation value of $X$ is that any such quantum state must necessarily have a large entropy in the computational basis\cite{Hastings_2018}.  Thus, not only must there exist computational basis states states with large $|H_Z|$, there must exist many such states.

\subsection{Dense Case}
A final interesting case to consider is a dense case, $N_T\sim N^K$.  
The dense case was studied previously\cite{haastad2004advantage} where it was shown that one can in general improve upon a random assignment by an amount proportional to $\sqrt{N_T}$.
This means that one can achieve $\langle H_Z \rangle \sim N^{K/2}$ in the worst case.  This is interesting as the problem has degree $D \sim N^{K-1}$ and so the improvement over random even in the worst case is by much more than $N_T/D$ for $K>2$.

In fact, the algorithm of Ref.~\cite{haastad2004advantage} is very simple, consisting simply of randomly sampling solutions until one achieve a solution with the given improvement.  Indeed, the fluctuations in the expectation value of $H_Z$ that we have written as $o(N)$ above simply reflect this variance in the solution.

However, it is interesting to analyze what happens with the quantum algorithm.  Consider the Hamiltonian
\be
H=X+\frac{1}{\sqrt{N_T}} H_Z,
\ee
which amounts to choosing $\alpha=D/\sqrt{N_T}$.

We analyze this Hamiltonian using a Krylov subspace: define the three states $$|0\rangle=|\pplus\rangle,$$ $$|1\rangle=c H_Z |0\rangle,$$
where the scalar $c=(\langle 0 | H_Z^2 |0\rangle)^{-1/2}=N_T^{-1/2}$ is chosen to normalize the state,
and
$$|2\rangle=d H_Z|1\rangle+e |0\rangle,$$
where the scalars $d,e$ are chosen to make $|2\rangle$ normalized and orthogonal to $|0\rangle,|1\rangle$.
Note that $|0\rangle$ and $|1\rangle$ are both eigenstates of $X$ so that the first three basis vectors of the Krylov subspace generated from $|0\rangle$ by $H$ are the same as those generated from $|0\rangle$ by $H_Z$.

Restricting the Hamiltonian $H$ to this three-dimensional subspace we have the tridiagonal Hamiltonian
$$
\begin{pmatrix}
N & H_{01} & 0\\
H_{01} & H_{11} & H_{12} \\
0 &H_{12} &H_{22}
\end{pmatrix}.$$
We have $H_{01}=1$.

The diagonal entry $H_{11}=N-K+\frac{1}{N_T^{3/2}} \langle \pplus | H_Z^3 | \pplus\rangle$ can be bounded by
$N_T^{3/2} 2^K$ by a theorem of Bonami\cite{bonami1970etude} that implies for any $H_Z$ that is a sum of terms of order at most $K$
that
\be
\label{Bonami}
|\langle \pplus | H_Z^p | \pplus \rangle| \leq (p-1)^{Kp/2} \langle \pplus | H_Z^2 | \pplus \rangle^{p/2}
\ee
for any $p\geq 2$.
Then, $|H_{11}-(N-K)| \leq O(2^{3K/2})$.

The state $|1\rangle$ is an eigenstate of $X$ with eigenvalue $N-K$.
So, the entry $H_{12}$ can be bounded by $\langle 1 | H_Z^2 | 1 \rangle$.
We have
$\langle 1 | H_Z^2 | 1 \rangle \leq  3^{2K}+N^2$,
using Eq.~(\ref{Bonami}).
So, $| H_{12}| \leq 3^{2K}$.

Since we have bounded terms $H_{01},H_{12}$ and bounded the difference $H_{11}-H_{00}$ where $H_{00}=N$, bounding all of these terms by quantities that are independent of $N$, and $O(3^{2K})$,
it follows that no approximate eigenstate,
can have almost all of its probability on state $|0\rangle$.
Precisely, let $\psi$ be a state such that $|H\psi-E\psi|\leq \exp(-O(K))$.
Then, $\psi$ cannot have more than $1-\exp(-O(K))$ of
its probability on state $0\rangle$.  Otherwise, $\langle 1 | (H-E) | \psi \rangle$ would be too large as the term $H_{01}=1$.

Hence, the state $\psi$ cannot have $\langle \psi | X | \psi \rangle \geq N-\exp(-O(K))$ and so it must have
$\langle \psi | H_Z | \psi \rangle \geq \frac{1}{\sqrt{N_T}} \exp(-O(K))$.

\section{Analysis of Quantum Algorithm}
\label{aqa}
We now analyze the quantum quench algorithm in more detail.
From Eq.~(\ref{energybal}),
$\langle H_Z\rangle_T =D \frac{N-\langle X\rangle_T}{\alpha}$.
Consider any given site $i$.  We will estimate $\langle X_i \rangle_T$.  Summing over $i$ will give
$\langle X \rangle_T$.

The basic physical idea is that if we can ignore the time-dependence of the force $F_i$, then we can approximate
$\langle X_i\rangle_T$ by the expectation value of $X_i$ assuming that the qubit $i$ evolves for a time $T$ under a time-independent Hamiltonian.
This time-independent Hamiltonian has a transverse field of strength $1$ (i.e., the term $X_i$ in the Hamiltonian) and a parallel field $(\alpha/D) F_i$, where $F_i$ is the force assuming that all other qubits $Z_j$ for $j\neq i$ are drawn from a uniformly random distribution (because at time $T=0$, the state of the system is $\pplus$ which has equal amplitude on all states).
In this case, similar to the analysis of the classical algorithm before, the force $F_i$ is likely to be at least of order $\sqrt{D}$ in which case we will have
$1-\langle X_i \rangle_Y \sim (\alpha/D)^2 \langle F_i^2\rangle_+ T^2 \sim \alpha^2 T^2/D$.

However, we cannot always neglect the time-dependence of the force.  To estimate whether or not the time-dependence of the force is important, we should compare the time-derivative of the force to $\sqrt{D}/T$.  If the time-derivative of the force is small enough compared to $\sqrt{D}/T$, then the approximation of the above paragraph will be valid.

In subsection \ref{tif} we analyze the time-independent case.
Subsection \ref{harmonosc} describes a toy example where we can see the effects of time-dependence.
In subsection \ref{tdf} we consider the time-dependence in more detail.

\subsection{Time-Independent Force}
\label{tif}
Let us first analyze the time-independent force approximation in more detail before considering the time-dependence.
We wish to compute
$$\langle \pplus| \exp(i (\frac{\alpha}{D}Z_i F_i+X_i)T)  X_i \exp(-i(\frac{\alpha}{D}Z_i F_i+X_i)T) |\pplus\rangle.$$  That is, we are considering an evolution under a Hamiltonian which includes the coupling $(\alpha/D) Z_i F_i$ and the transverse field $X_i$, but ignores any other coupling terms which would give the remaining qubits a time-dependence in the $Z$-basis.

As in the analysis of the classical case, the 
probability that $|F_i| \geq \sqrt{D}$ is at least
$\frac{1}{4}\exp(-4K)$.
At the same time, by Eq.~(\ref{boundeq}),
the probability that $|F_i| \geq t \sqrt{D}$ for $t\geq (2e)^{K/2}$ is at most
$\exp(-\frac{K}{2e} t^{2/K})$.
Picking $t$ sufficiently large (for example, $t=C^{K/2}$ for sufficiently large, $K$-independent constant $C$ suffices), this probability is much smaller than $\frac{1}{8} \exp(-4K)$.
So, with probability 
at least $\frac{1}{8}\exp(-4K)$, we have $|F_i|\in [\sqrt{D},C^{K/2} \sqrt{D}]$.

Then, for $C^{K/2}\sqrt{D} T$ sufficiently small compared to $1$, for any $|F_i|$ in that interval we have
\be
\langle \exp(-i (\frac{\alpha}{D}Z_i F_i+X_i)T) \pplus |X_i| \exp(-i(\frac{\alpha}{D}Z_i F_i+X_i)T) \pplus \rangle
 \leq
1-
\frac{\alpha^2 T^2}{D} \exp(-O(K)).
\ee

Thus, if this time-independent approximation is valid (and valid for all $i$) we have
that
\be
\langle H_Z \rangle_T \geq \alpha T^2 \exp(-O(K)) N.
\ee

Remark: here we required an upper bound on force $F_i$ because of the fixed time.  If we average over times on an interval, such an upper bound is not necessary.

\subsection{Toy Example}
\label{harmonosc}
The Hamiltonian of Eq.~(\ref{ex2}) provides an interesting example to study the time-dependence of the force.
Defining $Z=\sum_i Z_i$,
for the Hamiltonian (\ref{ex2}) we have $H_Z=-\frac{1}{2} Z^2 + {\rm const.}$  (This constant is positive and of order $N$.)
Hence, up to an additive constant, we have
$H=X-\frac{\alpha}{2D} Z^2\approx X-\frac{\alpha}{2N}Z^2$, since $D=N-1$.
This system can be approximately treated as a harmonic oscillator, at least for $X$ close to $N$.
We work in an eigenbasis of $Z$, letting state $|z\rangle$ denote an eigenstate of $Z$ with eigenvalue $z$.  In the large $X$ regime, the wavefunction has most of its probability on basis states with $i$ close to zero where the $X$ operator is approximately equal to $(N/2) |z\rangle\langle z+1| + h.c.$.  We can approximate further by treating $z$ as a continuous variable, approximating 
$(N/2) |z\rangle\langle z+1| + h.c.$ by $N+(N/2) \partial_z^2$, valid in the long wavelength regime.
We find then that the Hamiltonian is approximately equal to (ignoring additive constants) $$\frac{N}{2} \partial_z^2-\frac{\alpha}{2N}z^2.$$

Other than the overall sign, this Hamiltonian is the familiar Hamiltonian for a harmonic oscillator.
The oscillator has angular frequency
\be
\omega=\sqrt{\alpha}.
\ee

The $z$ variable oscillates periodically with time at the given frequency.  The force $F_i$ at time $t$ is (in this continuum approximation) equal to
$z(t)$.
Hence, if $\alpha T^2\gtrsim 1$, then the time-dependence of the force cannot be neglected in this example.  Note that here again we see this product $\alpha T^2$ appearing; in the time-independent analysis above (and in the previous heuristic analysis), this product controls the expectation value of $H_Z$.  Thus, it is no surprise that for this toy example
the time-independent approximation breaks down since there is no way to make the expectation value of $H_Z$ be large compared to $N$ for this instance.

\subsection{Time-Dependent Force}
\label{tdf}
We now consider the case of time-dependent force.
Intuitively, one may expect that for the time-independent approximation to break down, the magnitude
$\dot F_i$, i.e., ``how quickly the force is changing in time", must be comparable to the force at time $0$ (i.e., to $\sqrt{D}$) divided by the time $T$, in order for the force to be small at time $T$.
We will show this more precisely using Cauchy-Schwarz inequalities.

One might then guess (we do {\it not} show this) that given $\dot F_i$ of order $\sqrt{D}/T$, and given that $\langle 1-X_i \rangle_T$ is of order
$\alpha^2 T^2/D$, then $Y_i$ would be of order $\alpha T/\sqrt{D}$ and so $Y_i \dot F_i$ would be of order $\alpha$.
Thus, the results in this section may be interpreted as evidence in favor of conjecture \ref{conjer}.
(Note that if $\langle 1-X_i \rangle_T$ is much larger than this, then Eq.~(\ref{energybal}) guarantees a large expectation value for $H_Z$ while if $\langle 1-X_i \rangle_T$ is much smaller than this, the constraints on the state at time $T$ become more stringent due to the larger expectation value for $X$.  Further, the magnitude of $\dot F_i$ would need to be larger to have a smaller expectation value $\langle 1-X_i \rangle_T$.)

Define
\be
\Delta_s=F_i-\tau_s^H(F_i).
\ee
Define
\be
\phi(T)=\exp(-i T H)
\exp\Bigl(-i\frac{\alpha}{D}\int_0^T Z_i \Delta_s {\rm d}s\Bigr) \pplus.
\ee
This state $\phi(T)$ has the following property as can be seen by going to the interaction representation.
Define the operator $R$ by
\be
H=X_i+Z_i F_i + R,
\ee
so that $R$ includes all terms in $H$ which are not supported on site $i$.
Then $\phi(T)=\exp(-iRT) \exp(-i(\frac{\alpha}{D}Z_i F_i+X_i)T) \pplus$.
Then,
\be
\langle \phi(T) | X_i | \phi(T)\rangle =  \langle\pplus| \exp(i (\frac{\alpha}{D}Z_i F_i+X_i)T)X_i \exp(-i(\frac{\alpha}{D}Z_i F_i+X_i)T) |\pplus\rangle,
\ee
so that the expectation value 
$\langle \phi(T) | X_i | \phi(T)\rangle$ is given by the time-independent approximation above.

We have 
\begin{eqnarray}
\phi(T)&=&
\exp(-i T H)
\exp\Bigl(-i\frac{\alpha}{D}\int_0^T Z_i \Delta_s {\rm d}s\Bigr) \pplus
\\ \nonumber
&=& 
\exp(-iTH) \pplus +\xi,
\end{eqnarray}
where the exponential is an $s$-ordered exponential (i.e., it is time-ordered with respected to $s$, as are later exponentials of integrals below)
and where we define
\be
\xi =-i\frac{\alpha}{D}\exp(-i T H)\int_0^T {\rm d}s \,
\exp\Bigl(-i\frac{\alpha}{D}\int_s^T \Delta_u {\rm d}u\Bigr) Z_i \Delta_s
\pplus.
\ee
So,
since $\exp(-i\frac{\alpha}{D}\int_s^T \Delta_u {\rm d}u)$ is unitary, by a triangle inequality we have
\be
\label{xibound}
|\xi|\leq \frac{\alpha}{D}\int_0^T {\rm d} s \, \sqrt{\langle \Delta_s^2 \rangle_+}.
\ee

Define
\be
\pplus(T)=\exp(-i TH) \pplus.
\ee
So,
\be
\phi(T)=
\pplus(T)+\xi.
\ee

Hence,
\be
\langle \phi(T) | X_i | \phi(T) \rangle=\langle \tau_T(X_i) \rangle_+ + 2 {\rm Re}
\langle \pplus(T) | X_i | \xi \rangle  +\langle \xi | X_i | \xi \rangle.
\ee
Let $\Pi^-_i=(1-X_i)/2$, so that it projects onto the $|-\rangle$ state on qubit $i$.
So,
\be
\langle \phi(T) | \Pi^-_i | \phi(T) \rangle=\langle \tau_T(\Pi^-_i) \rangle_+ + 2 {\rm Re}
\langle \pplus(T) | \Pi^-_i | \xi \rangle + \langle \xi | \Pi^-_i | \xi \rangle.
\ee
By Cauchy-Schwarz, the second term in the above equation is bounded by $2 \sqrt{\langle \tau_T^H(\Pi^-_i) \rangle_+} |\xi|.$
The third term is bounded by $|\xi|^2$.

Hence,
\be
\langle \phi(T) | \Pi^-_i | \phi(T) \rangle\leq \langle \tau_T(\Pi^-_i) \rangle_+ + 2 \sqrt{\langle \tau_T^H(\Pi^-_i) \rangle_+} |\xi|
+|\xi|^2.
\ee
So,
\be
\langle \tau_T^H(\Pi^-_i) \rangle_+\geq \langle \phi(T) | \Pi^-_i | \phi(T) \rangle-2 \sqrt{\langle \tau_T^H(\Pi^-_i) \rangle_+} |\xi|-|\xi|^2.
\ee
Thus,
\be
\langle \tau_T^H(\Pi^-_i) \rangle_+\geq \langle \phi(T) | \Pi^-_i | \phi(T) \rangle-2 \sqrt{ \langle \phi(T) | \Pi^-_i | \phi(T) \rangle} |\xi|-|\xi|^2.
\ee

So, if we can bound $|\xi|$ sufficiently small compared to $\sqrt{\langle \phi(T) | \Pi^-_i | \phi(T) \rangle}$, then we lower bound 
$\langle \tau_T^H(\Pi^-_i) \rangle_+$ compared to  $\langle \phi(T) | \Pi^-_i | \phi(T) \rangle$.
For example,
 if we can bound that $|\xi|\leq  \sqrt{\langle \phi(T) | \Pi^-_i | \phi(T) \rangle}/3$, then
$\langle \tau_T^H(\Pi^-_i) \rangle_+\geq \langle \phi(T) | \Pi^-_i | \phi(T) \rangle \cdot (1-2/3-1/9)=(2/9) \cdot \langle \phi(T) | \Pi^-_i | \phi(T) \rangle.$
If we can give even tighter bounds on $|\xi|$, then $\langle \tau_T^H(\Pi^-_i) \rangle_+\rightarrow \langle \phi(T) | \Pi^-_i | \phi(T) \rangle$ as $|\xi|\rightarrow 0$.

So, we now bound
$|\xi|^2$.  From Eq.~(\ref{xibound}),
$|\xi|\leq \frac{\alpha}{D}\int_0^T {\rm d} s \, \sqrt{\langle \Delta_s^2 \rangle_+}$.
So, we turn to bounding $\langle \Delta_s^2 \rangle_+$.
We have
$\Delta_s \pplus=-\int_0^s {\rm d}v \, \tau_v^H(\dot F_i),$ since $\Delta_0=0$.
So, again by Cauchy-Schwarz,
\be
\langle \Delta_s^2 \rangle_+ \leq s \int_0^s {\rm d}v \, \langle \Bigl| \tau_v^H(\dot F_i)\Bigr|^2 \rangle_+.
\ee
So,
\be
|\xi| \leq \frac{\alpha}{D} \int_0^T {\rm d}s \, s \sqrt{\frac{\int_0^s {\rm d}v \, \langle \Bigl| \tau_v^H(\dot F_i)\Bigr|^2 \rangle_+}{s}}.
\ee
So, 
$|\xi|$ is bounded by $\alpha T^2/(2D)$ times the expectation value of $\sqrt{\langle \Bigl| \tau_v^H(\dot F_i)\Bigr|^2 \rangle_+}$
for $s$ randomly chosen in the interval $[0,T]$ from measure $(T^2/2)^{-1} s {\rm d}s$ and $v$ uniformly randomly chosen in the interval $[0,s]$
This random choice of $s$ followed by a random choice of $v$ induces a measure
\be
{\rm d}\mu(v)=2(1-v){\rm d}v.
\ee

Thus, to have 
$\langle \tau_T^H(\Pi^-_i) \rangle_+/\langle \phi(T) | \Pi^-_i | \phi(T) \rangle$ small compared to $1$, given that $\langle \phi(T) | \Pi^-_i | \phi(T) \rangle$
is at least $\exp(-O(K)) (\alpha^2 T^2)/D$ as shown in subsection \ref{tif}, then we need that for random choice of $v$ from the measure
$\mu(v)$, that
the expectation value
$\expec_v[\sqrt{\langle \Bigl| \tau_v^H(\dot F_i)\Bigr|^2 \rangle_+}]$ is
at least
\be
\label{velocity}
\exp(-O(K)) \sqrt{\frac{\alpha^2 T^2}{D}} \frac{2D}{\alpha T^2} = \exp(-O(K)) \frac{\sqrt{D}}{T}.
\ee
This gives the intuitive result for the magnitude of $\dot F_i$ mentioned at the start of this subsection.

\section{Discussion}
We have considered an algorithm that uses quantum quenches as well as a classical algorithm to perform approximate optimization.  We have been shown (using $\epsilon$ slightly larger than $1/\sqrt{D}$ in theorem \ref{dualclemma}) that the classical algorithm improves upon random by a factor that is more than $1/D$ {\it unless} the problem has a ``very bad" solution, i.e., one that is $\Omega(1)$ worse than random.  This can be used then in some cases to guarantee that the algorithm will find a nontrivial improvement if no such very bad solution exists.  We have also given a heuristic analysis of the quantum algorithm

We emphasize that this quench algorithm is not described by a fixed depth quantum circuit, independent of $D$.  The Lieb-Robinson velocity $v_{LR}$ of
this Hamiltonian is proportional to $\sqrt{\alpha}$, as can be shown by using Lieb-Robinson bounds adapted to Hamiltonians which are a sum of two types of terms (in this case, $X_j$ for different qubits $j$ is one type and terms in $H_Z$ is another type) such that terms within a type commute\cite{premont2010lieb}; more generally, we can use bounds adapted to the case of a bounded commutator\cite{haah2018quantum}.  Here to define the Lieb-Robinson velocity, we define a distance between qubits by using a graph metric for
a graph with vertices corresponding to qubits and an edge between vertices if the corresponding qubits are both in some term in $H_Z$.  

The estimates using the Lieb-Robinson velocity give
some upper bound on how far a perturbation can propagate in a given time; the effect of a perturbation beyond a distance proportional to $v_{LR} t$ is negligible.  These estimates may not be tight, but we expect that indeed the
velocity of perturbations will be proportional to $\sqrt{\alpha}$ in many systems.  If this is true, then if $\alpha t^2$ diverges with $D$ to obtain a nontrivial approximation,
the necessary circuit depth also diverges.

Also, it may be useful to consider a generalization of the algorithm in which one does some slow (but not necessarily adiabatic) evolution of the Hamiltonian from an initial Hamiltonian $X$ to $H=X+(\alpha/D) H_Z$, followed by an additional time evolving under $H=X+(\alpha/D) H_Z$.
This is similar to the quantum adiabatic algorithm except on proceeds at some nonzero speed, allowing level crossings.  The point of the analysis here is that even if the evolution from initial to final Hamiltonian is very nonadiabatic, the evolution for a nonzero time in some fixed Hamiltonian can achieve a useful result as decohering in the eigenbasis can increase the expectation value of $H_Z$ while reducing that of $X$.
This decoherence is a possible principle that can be used to show a nontrivial approximation.

.

\bibliography{qqn-ref}

\end{document}